\documentclass[conference,letterpaper]{IEEEtran}
\IEEEoverridecommandlockouts
\usepackage{cite}
\usepackage{amsmath,amssymb,amsfonts}
\usepackage{amsmath,amssymb,amsfonts}
\usepackage{textcomp}
\usepackage{comment}
\usepackage{algorithm2e}
\usepackage{graphicx}
 \usepackage[numbers]{natbib}
 \usepackage{soul}
\usepackage{color}
\usepackage{flushend}
\usepackage{amsmath,amssymb,amsfonts}
\usepackage{algorithmic}
\usepackage{graphicx}
\usepackage{textcomp}
\usepackage[OT1]{fontenc}
\usepackage{pifont}
\usepackage{multirow}
\usepackage{threeparttable}
\usepackage{booktabs}

\usepackage[dvipsnames,svgnames]{xcolor, colortbl}

\definecolor{LightCyan}{rgb}{0.88,1,1}

\newcommand\blfootnote[1]{%
  \begingroup
  \renewcommand\thefootnote{}\footnote{#1}%
  \addtocounter{footnote}{-1}%
  \endgroup
}

\def\BibTeX{{\rm B\kern-.05em{\sc i\kern-.025em b}\kern-.08em
    T\kern-.1667em\lower.7ex\hbox{E}\kern-.125emX}}
\begin{document}
\bstctlcite{IEEEexample:BSTcontrol}

\title{Behavior-neutral Smart Charging of Plugin Electric Vehicles: Reinforcement learning approach}
\author{
\IEEEauthorblockN{
Vladimir Dyo
}
\IEEEauthorblockA{
\textit{School of Computer Science and Technology, University of Bedfordshire, UK }\\
vladimir.dyo@beds.ac.uk}
}

\maketitle
\begin{abstract}
High-powered electric vehicle (EV) charging can significantly increase charging costs due to peak-demand charges. 
This paper proposes a novel charging algorithm which exploits typically long plugin sessions for domestic chargers and reduces the overall charging power by boost charging the EV for a short duration, followed by low-power  charging for the rest of the plugin session. 
The optimal parameters for boost and low-power charging phases are obtained using reinforcement learning by training on EV's past charging sessions. 
Compared to some prior work, the proposed algorithm does not attempt to predict the plugin session duration, which can be difficult to accurately predict in practice due to the nature of human behavior, as shown in the analysis.  
Instead, the charging parameters are controlled directly and are adapted transparently to the user's charging behavior over time. 
The performance evaluation on a UK dataset of 3.1 million charging sessions from 22,731 domestic charge stations, demonstrates that the proposed algorithm results in 31\% of aggregate peak reduction. 
The experiments also demonstrate the impact of history size on  learning behavior and conclude with a case study by applying the algorithm to a specific charge point. 
\end{abstract}


\section{Introduction}
\label{secIntro}
\blfootnote{Accepted to IEEE Access, vol. 10, pp. 64095-64104, 2022, doi: 10.1109/ACCESS.2022.3183795.}
As electric vehicles become more popular due to higher energy efficiency and lower running costs,
additional power demand required for charging becomes a key challenge. 
In the UK, electric vehicles could add 24\,Gwatt of additional peak electricity demand by 2050, an extra 30\% of current capacity \cite{nationalGridFES}.
Managing the peak power demand, therefore, becomes critically important  to protect the grid from overload while supporting the uptake of electric vehicles \cite{energyTaskForce}. 

Smart charging is  critical to reducing peak demand and operates by time-shifting charging to whenever the grid has spare capacity and may incentivize users to change their charging patterns through dynamic pricing. 
The concept has been explored for  charging individual vehicles  and entire fleets  \cite{Vandael2015smartgrid}.  
For example, Tomic and Kempton \cite{tomic2007powersources} study the 
 impact of time-based charging, which schedules charging to off-peak hours. 
Lacey et al \cite{lacey2017iet} compare the impact of uncontrolled and smart charging strategies on local grid load.

However, smart charging often  requires  cooperation from the driver, e.g. a willingness to  postpone charging to off-peak hours.  
On the other hand, the adoption of dynamic pricing depends on whether users {\em en masse} are willing to change their habits in exchange for lower energy prices \cite{energyTaskForce}. 
At the moment, there is evidence that many users choose to charge at standard rates whenever they require to charge their EVs, which could be related to range anxiety as drivers want to make sure their EVs are fully charged whenever possible. 
In addition, in home-based scenarios, the dynamic pricing is too complex for an individual to process to make an informed decision \cite{LATINOPOULOS2017175}.  

This paper explores a novel  charging algorithm  that learns the users' personal charging history and optimizes the charging current to satisfy the energy demand without cooperation from the driver. 
The study first analyzes the predictability of the plugin session durations and demonstrates that in the absence of additional contextual information, the duration of the plugin session cannot be accurately predicted, and therefore cannot be relied upon to spread the charging power evenly through the charging session.  
 Then a reinforcement learning-based approach is developed, which analyzes the past charging history to directly regulate the charging current and reduce the peak demand. 

The proposed approach operates by boost-charging the EV at full speed for a limited duration of time and then slow charging at a fraction of a full speed until the EV  charge is complete. 
The evaluation using a UK dataset of 3.2 million domestic EV charge sessions from 22,731 domestic charge points shows 31\% of aggregate peak reduction. 
The performance is also compared with  uncontrolled charging, where a vehicle is charged at maximum rated power until charged, and with hypothetical oracle-based charging, where power is spread evenly throughout a plugin session.

The paper is structured as follows:
Section \ref{secRelated} describes related work.  
A prediction model based on linear regression is described in Section \ref{secPrediction}. 
Section \ref{secAdaptive} describes reinforcement learning-based approach. 
The details on the dataset and statistical analysis are presented in Section \ref{secDataset}.
Section \ref{secResults} contains experimental results and discussion followed by Conclusion in Section \ref{secConclusion}.

\section{Related work}
\label{secRelated}

It is well known that uncontrolled charging of EVs can lead to power outages, reduce power quality and increase power losses and operating costs. From the user's perspective, however, uncontrolled charging is the most natural, as he or she can charge the EV wherever and whenever needed. 
Smart charging attempts to match grid capability with EV energy demand by actively controlling EV charging parameters such as output power, charging time, duration, and time pattern. As charging parameters change, the user must adapt its behavior or face higher charging costs. 

Centralized strategies compute a globally optimal solution based on information collected from individual EV vehicles, grid conditions, market information, and statistical  forecasting models. 
The central controller, also known as an aggregator, collects data from EVs such as state of charge and user preferences and then based on the selective objective applies an algorithm to compute setpoints for individual EVs. 
Depending on the objective, smart charging can maximize operator’s profit \cite{Han2010smartgrid}, minimize energy costs \cite{wu2012smartgrid}\cite{Vandael2015smartgrid}, maximize EV utility, or ensure fairness. 
This strategy, however, requires expensive communication infrastructure including standardized architecture and protocols, which are still under development. 
Centralized charging is suitable for charging fleets of vehicles such as buses or utility vehicles, which tend to have more predictable energy consumption patterns.  

In decentralized strategies, the decision making is done locally using price signals and driver preferences and can be done autonomously or in cooperation with other users or energy controllers in the area. 
The decisions range from postponing charging to off-peak hours \cite{lacey2017iet} to adjusting the charging rate adaptively depending on the dynamic pricing. 
The former strategies are simpler but may result in an avalanche effect, when many EVs simultaneously select a similar action, e.g., postpone charging to off-peak hours, which may create an unwanted power consumption spike. 
Whereas the dynamic pricing-based strategies would rely on the users changing their charging behavior and pattern to minimize their energy costs.

The early works on EV charging used simulations to model the charging load, and relied on various assumptions to produce tractable models. 
Kelly et al \cite{Kelly2009epec} develop a probabilistic model based on Monte Carlo simulations to produce load profiles (including uncontrolled EV charging) for residences, offices, and retail stores.  
For office and retail locations, the model assumes a fixed charging rate, an EV arrival within a certain time interval, and charging from  a random battery state until the battery is full. 
The simulations for retail locations use traffic volume data to compute the probability of vehicle arrival during each 30-min slot throughout a day. 
For residential PEVs, the model assumes that vehicle owners commute every day and uses statistical data from US National Personal Transportation Survey. 
Probabilistic approaches often rely on assumptions to produce more tractable models.  
For example, Soares et al \cite{Soares2011ASM} assume that all parked vehicles are connected for charging, which may not be realistic. 
Steen et al \cite{steen2012smartgrid} use demographic data such as EV locations, number of workspaces and employees, and usage patterns from national travel surveys to estimate charging behavior and control charging. 
Shahidinejad et al \cite{Shahidinejad2012smartgrid} introduce a subjective decision making process of whether  a driver plugs in a parked vehicle  for charging, to compute aggregate charging load more accurately, whereas
\cite{Csaba2011impact}\cite{munoz2012fuzzyimpact}\cite{Zheng2013smartgrid}  use the Poisson arrival process to model the number of vehicles arriving at a certain location  to compute an aggregate charging load from multiple vehicles.

More recently, Lacey et al \cite{lacey2017iet}  evaluated  smart charging strategies using realistic non-charging electricity load profiles, which are used in the design of electricity networks and show the aggregated demand over a large number of users over time \cite{ElNozahy2014phev}, and overlaying them with an EV charger load. 
It should be noted that the electricity load profiles typically have seasonal variation and are different for domestic and industrial users. 
\cite{wei2018ieeetrans} developed an algorithm to reduce  the cost of EV battery degradation and the peak power load but evaluated it in simulation using synthetic datasets  without using any real datasets. 
\cite{Kisacikoglu2018ieeetrans} proposed a distributed smart charging algorithm to reduce peak power at a charging site while satisfying each EV energy demand but  evaluated it using a generated mobility dataset that contains daily distances, home arrival, and departure times using predefined Gaussian characteristics.

\subsection{Data-driven approaches}
The emergence of massive datasets has enabled data-driven approaches, which allow evaluating the system design directly on actual usage data \cite{zhang2018smartcharging}\cite{Fenner20peakshaving}\cite{Xydas2016datadriven}\cite{ali2020isgt}\cite{Buzna2019synergy}. 
A large proportion of works using real datasets focused on characterisation of demand \cite{Xydas2016datadriven}, studying charging behavior \cite{Wolbertus2016mdpi}, prediction \cite{Pevec2018energyresearch} and other \cite{shipman2019energy}\cite{winsys17}  aspects.

\subsubsection{Smart charging}

The availability of real charging session data has stimulated the development of novel smart charging approaches. 
Zhang et al \cite{zhang2018smartcharging} designed a real-time algorithm for peak demand reductions at non-residential charging sites.
The approach achieves up to 80\% demand reduction, however, does not attempt to avoid a reduction in  the quality of charging service. 
Fenner et al \cite{Fenner20peakshaving}  investigated the maximum possible peak demand reduction capacity and conducted a case study in Finland by applying  various optimization strategies to  real data from 25,000 charging sessions collected over 2 years  from 8 charging sites, and  show that the peak loads at charging sites can be reduced by up to 55\%.  
However, the optimization strategy used in the study  computes  the peak load  as a ratio of dispensed energy to plugin duration, which requires the knowledge of the latter and corresponds to the hypothetical scenario in the presented study.

\subsubsection{Charging behaviour analysis}

Xydas et al \cite{Xydas2016datadriven}   develop a fuzzy-logic-based model to characterize EV charging demand depending on weather and trend. 
The approach estimates the monthly growth rate of EV charging demand using linear regression and measures the correlation between weather attributes and the daily peak power of EVs charging in a geographical area. 
The output is then used by fuzzy-logic-based module to establish the level of risk to grid operation using a dataset containing 21,918 EV charging events from 255 charging stations  in the UK for evaluation. 
Although the  authors classify households with EVs based on their energy usage patterns, there is no attempt to predict the EV energy demand or availability at the individual charger level. 

Wolbertus et al \cite{Wolbertus2016mdpi} study the charging infrastructure utilization in 5 cities in the Netherlands based on 1.6 million charge sessions  from 5,600 charge points over  two years. 
The authors aim to identify different charge patterns and charge behavior depending on the area. 
Hence the analysis is done on an aggregate rather than individual charger level. 
Similarly, Buzna et al \cite{Buzna2019synergy} analyze the aggregate load from EV charge stations  using machine learning and time-series analysis and evaluate the approach on EVnetNL dataset from the Netherlands, which contains over 32 million sessions from over 1,700 charge points. 
Straka et al \cite{Straka2020access} developed a method for predicting the popularity of EV charging infrastructure using EVnetNL dataset  in combination with GIS data. 
The approach predicts whether a given charge spot belongs to a top tier using binary classification and logistic regression. 
Finally, Pevec et al \cite{Pevec2018energyresearch} propose a methodology to combine multiple data sources, including places of interest near chargers, the driving distance between the chargers, and historical data about charging transactions to predict charging station utilization when the contextual data change or when there is a change in charging infrastructure.

\section{Session Predictability}
\label{secPrediction}

\label{secTechnology}

This section analyses  the predictability of plugin session duration based on the history of past plugin sessions and shows the overall prediction accuracy together with the impact of each feature on the overall prediction accuracy. 
The motivation for predicting the plugin session duration is that it can be used by the EV charger to evenly spread the required energy to  reduce the peak demand. 

The following features were extracted from the dataset to predict the current charging session duration: 
 session start hour, day of the week, time duration since the last charging session, and the amount of dispensed energy. 
Although the dataset shows the amount of dispensed energy for each session, it  does not indicate whether the EV was fully charged or whether it was just a top-up. 
In the former case, the amount of dispensed energy can be assumed to be known at the start of the charging session. 
In the latter, i.e. a top-up session, the amount of dispensed energy becomes known at the end of the charging session for which the duration needs to be predicted.
To account for both cases,  the regression performance was conducted with and without this feature. 

\subsection{Linear regression}
Linear regression models a continuous variable $y_j$ as a linear combination of independent variables $X$. 
The advantage of regression analysis methods is that they are computationally efficient and are simple to understand. 
\begin{equation}
y_j = \beta_0 + \sum_{i=1}^t\beta_i x_i + \epsilon_j
\end{equation}
Where $\beta_0$ is an intercept, $\beta_i$ is a slope, $t$ is the number of observations, $\epsilon_j$ is an error term, the part of the model that cannot explain the linear relationship. 
The regressor weights are obtained during the training phase as  the ratio of covariance between  $x_i$ and  $x_j$ and the variance  of $x_i$:
\begin{equation}
\beta_i = \frac{cov(x_i, x_j)}{var(x_i)}
\end{equation}

\subsection{Performance Metrics}
The regression performance has been evaluated with mean absolute error (MAE), mean absolute percentage error (MAPE) and mean square error (MSE) metrics defined as shown below for reference. 
\begin{equation}
\footnotesize
MAE = \sum_{n=1}^N\frac{|predicted - actual |}{N}
\end{equation}

\begin{equation}
\footnotesize
MAPE = \sum_{n=1}^N\frac{|predicted - actual |}{actual}\times \frac{100\%}{N}
\end{equation}

\begin{equation}
\footnotesize
MSE = \sum_{n=1}^N\frac{(predicted - actual)^2}{N}
\end{equation}

\subsection{Prediction accuracy}

\begin{table}[h!]
\smaller
\centering
\caption{Predicting session duration. }
\begin{tabular}{||c c c||} 
 \hline
 MAE & MAPE & MSE  \\ [0.5ex] 
 \hline\hline
 14.04 & 413.93 & 11517.59  \\ 
 \hline
\end{tabular}
\label{table:1}
\end{table}

The prediction accuracy was evaluated on each charge point using 4-fold cross-validation, separately for each charge point with the  total prediction accuracy  computed as an average for all charge points. 
The data analysis has been performed using $R$ statistical package \cite{R}. 
The overall prediction accuracy is 14.04 MAE and 413.93 MAPE. 
Upon close inspection, the high MAPE values are contributed by a number of sessions, where the session duration was significantly overestimated. 
While underestimating session duration is not critical and may result in supplying the target energy while reducing the load, overestimating the session duration is obviously detrimental to any predictive charging strategy. 
This is because attempting to spread energy for a longer time than the actual session duration will result in missing the energy target. 

The reason for low predictability is not in the limitations of the selected method, as similar results have been obtained using a variety of other techniques including deep neural network algorithms. 
The  latter  required an immensely higher amount of computational power but resulted in only a modest improvement in accuracy. 
The key reason is that the session duration is tightly related to human behavior, which is inherently hard to predict. 
A weather condition, a traffic jam, or a road accident,  personal plans  are as likely to affect a session duration as the past  history of charging sessions. 
Possibly, enriching the  data set with additional sensor data, such as weather, traffic conditions, or home occupancy sensors may improve the prediction accuracy. 
However, the conclusion from the experiments in this study is that given the history of plugin sessions  alone,  prediction accuracy is too low   for adaptive charging purpose.

\section{Reinforcement learning based Adaptive charging}
\label{secAdaptive}

 \begin{figure}[!h]
 \centering
  \includegraphics[width=5cm]{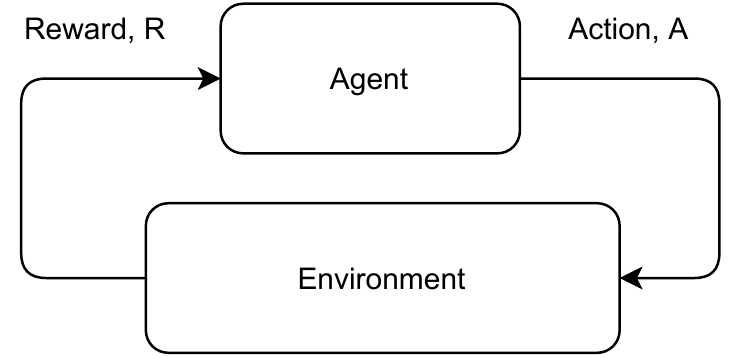}
  \caption{Reinforcement Learning approach. The agent interacts with the environment by selecting an action $A$ and receives a reward $R$. The goal of the agent is to select a policy that maximizes the reward. } 
  \label{fig:reinforcement}
\end{figure}

The reinforcement learning concept is based on the idea of an agent interacting with an environment and receiving the reward depending on the selected actions, Fig. \ref{fig:reinforcement}.
The agent learns the  policy that maximizes the reward through trial and error, by selecting various actions and observing the corresponding reward from the environment. 
The advantage of the reinforcement learning approach is that it  directly controls  the process in an uncertain environment without having to make any predictions about the environment itself. 

In the proposed approach, the reward is a function that is inversely proportional to the effective charging rate $P_{eff}$ and the energy loss as a result of applying a policy. 
The reinforcement learning algorithm  maximizes the reward by learning the optimal  charging parameters that minimize the effective charging rate while satisfying the energy demand. 
The agent is trained  using the history of past charging events and  applies the optimal parameters toward the next charging session, after which the  session becomes a part of the training dataset. 

 \begin{figure}[!h]
 \centering
  \includegraphics[width=7cm]{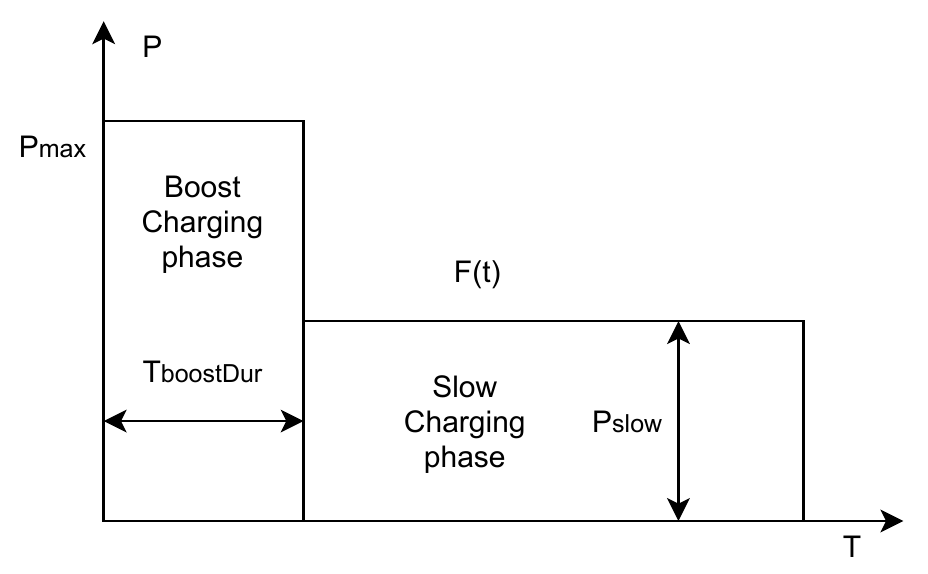}
  \caption{Charging function. The charging function consists of a boost charging and low-power (slow) charging phases and is characterized by parameters $T_{boostDur}$ and $P_{slow}$.  } 
  \label{fig:ChargingFunction}
\end{figure}

\subsection{Charging Function}
 
A charging function $F(t)$ is defined here as the target power profile for a given charging session over its entire duration, such that the total area under the curve is equal to a target amount of dispensed energy $\int F(t) = E_i$. 
 In the simplest case, a charging function $F(t)$ will have a constant value to represent charging at a certain power rate, and the learning algorithm would try to find an optimal value of the rate. 

In this study, a more complex charging strategy is considered, where a charging session consists of a boost charging phase at a maximum rate  $P_{max}$ for a duration of $T_{boostDur}$ followed by  low-power charging at a low power rate $P_{slow}$ for the rest of the session. 
The strategy can also be useful in energy management in battery-assisted charging systems, which accumulate energy in between charging sessions and then use it to boost-charge the EV. 

The selected charging function is defined by 2 parameters, $T_{boostDur}$ and a low power charge rate  
$P_{slow} = P_{optimal}P_{max}$. 
The parameters of $T_{BoostDur}$ and the coefficient $P_{optimal} \le 1$   are learned by an algorithm based on historical usage data and are updated dynamically after each charging event as described in the following subsection. 
It should be noted that throughout the paper the terms 'low power' and 'slow' charging will be used interchangeably.

\subsection{Reward function}
The agent's reward  is designed to decrease with either an energy loss $E_{loss}$ or  the aggregate charging rate $P_{aggr}$ over previous sessions:
\begin{equation}
R = 
\begin{cases}
-k_1 E_{loss} + k_2 /P_{aggr}, & if E_{loss} < E_{maxLoss}\\
 -\infty, & if E_{loss} \ge E_{maxLoss} 
\end{cases}
\end{equation}
To define a maximum acceptable energy loss, the reward function is set to negative infinity, if the energy loss exceeds a  certain threshold, which in this study was  selected as  $E_{maxLoss}$ = 10$\,$kWh. 
The  parameters $k_1$ and $k_2$ are constants that define the relative weights of energy loss and charging rate respectively. 
The optimal policy is such that minimizes the aggregate charging rate across all charging sessions in the past while ensuring that the energy loss is kept below a threshold. 

An energy loss  $E_{loss}$ is computed as a total sum of differences between the target $E_{target,i}$ and actually delivered energy amounts $E_{total, i}$ across all   past sessions:
$E_{loss} = \sum_i^N(E_{target,i} - E_{total, i})$,
 where $E_{target,i}$ represents the target amount of energy required  to charge the vehicle in session $i$ and  and is taken from the dataset. 
The actual  dispensed energy $E_{total, i}$ is always less than or equal than the target amount of energy, $E_{total, i} \leq E_{target,i}$ due to lower effective charging speed in adaptive mode.
The analysis shows that in the vast majority of domestic charging sessions the ratio of dispensed energy to plugin duration is lower than the maximum charging speed. 

The second component of the reward function, the aggregate power rate $P_{aggr}$ for each charge station, is computed as a sum of effective charging session rates weighted by the corresponding amount of dispensed energy:
\begin{equation}
\footnotesize
P_{aggr} = \sum{(P_{eff_i} E_{total,i} )} / \sum{E_{total,i}}
\end{equation}
The session effective charging rate $P_{eff_i} $ is computed as the charging speed weighted by the amount of energy dispensed at that speed:
\begin{equation}
\footnotesize
P_{eff, i} =  \frac{( E_{Boost, i}+  P_{rate}  (E_{Adap, i}-E_{Boost, i}))}   {E_{target, i}} P_{max}
\end{equation}
Where $E_{Boost, i}$ and $(E_{Adap, i} - E_{Boost, i})$ are the  amounts of energy dispensed in boost and low-power (slow) charge modes respectively. 
$P_{rate} \le 1$ is a candidate value of the slow charge rate coefficient, which is defined as  the proportion of the maximum power rate, $P_{max}$. 
Finally, $E_{Adap, i} \le E_{target}$ is the actual amount of energy delivered in adaptive mode respectively.  
The  session effective charging rate reduces with lower boost energy $ E_{Boost, i}$ and lower low-power charge rate $P_{rate}$, so the learning algorithm seeks to reduce those parameters as discussed in the next subsection.

\subsection{Training}
\begin{algorithm}[t]
\footnotesize
\DontPrintSemicolon
    \SetKwFunction{FMain}{LearnRLModel}

        \FMain{}{\\
  // {\em init the optimal boost duration} \\
  $T_{BoostDur} = T_{mean}$\\
  // {\em init optimal slow charge rate} \\
  $P_{optimal}$ = 0.5\\
  // {\em init reward value} \\
  $R_{optimal} = 0$\\
 \For{$(i \; in\; 1..n_{tries})$}{
  // {\em compute random step in boost phase duration} \\
 $\Delta X_i  = T_{mean} \times rand(\Delta X_{min}, \Delta X_{max})$\\
   // {\em compute new candidate value of boost phase duration} \\
 $T_{maxboost} = min(T_{BoostDur} \pm \Delta X_i, T_{plugin})$\\
 $T_{maxboost} = max(0, T_{maxboost})$\\
  // {\em compute random step in slow charge rate coefficient} \\
 $\Delta Y_i = rand(\Delta Y_{min}, \Delta Y_{max})$\\
   // {\em compute new slow charge rate} \\
$P_{rate} = min(P_{optimal} \pm \Delta Y_i, 1.0)$\\
$P_{rate} = max(0, P_{rate})$\\
  // {\em evaluate new candidate values } \\
($E_{loss}, P_{aggr}) = evaluateRLModel(data, T_{maxboost}, P_{rate}$)\\
  // {\em compute the reward} \\
\If{$E_{loss} > E_{maxLoss}$} {$R = -\infty$;} 
\Else {$R  = -k_{1}E_{loss} + k_2/P_{aggr}$}
\If{$R > R_{optimal}$} {
      $T_{BoostDur}$ = $T_{maxboost}$;\\
       $P_{optimal} = P_{rate}$;\\
      $R_{optimal} = R$;\\
    }  
 } 
$return(T_{BoostDur}, P_{optimal}$)
}

$\,$
\label{algo}
\caption{Adaptive charging  algorithm pseudocode}
\end{algorithm}

\begin{algorithm}[]
\footnotesize
\DontPrintSemicolon
    \SetKwFunction{FMain}{EvaluateRLModel}
        \FMain{data, $T_{maxboost}$, $P_{rate}$}{

  $P_{effective}$ = 0\\
 \For{(i in 1:$n_{tries}$)}{
  //{\em the actual amount of time in boost mode:}\\
  $T_{boost,i} = min(E_{target}/P_{max}, T_{maxboost}$)\\
  //{\em the amount energy dispensed in boost phase mode:}\\
  $E_{boost,i} = min(T_{boost,i} P_{max}, E_i$)\\
  //{\em the total amount of energy dispensed in adaptive mode:}\\
  $E_{total,i} = P_{max} (T_{boost,i} + (T_{plugin} - T_{boost}) P_{rate})$\\
  $E_{total,i} = min( E_{target}, E_{total,i})$\\
  //{\em the amount of energy dispensed in slow charge mode:}\\
  $E_{slow,i} = E_{total, i}  - E_{boost, i}$\\
  //{\em the duration of slow charge mode:}\\
  $T_{slow,i} =  E_{slow}/(P_{max}P_{rate})$\\   
  //{\em the effective charge rate:} \\
  $P_{eff, i} =  ( E_{Boost, i}+  P_{rate}  (E_{Adap, i}-E_{Boost, i}))\frac{P_{max}}{E_i}$\\

}
//{\em the total energy deficit:}\\
$E_{loss} = \sum(E_{target,i} - E_{total, i})$\\
//{\em the aggregate historical charge rate under given policy:}\\
$P_{aggr} = \sum{(P_{eff, i} E_{total,i} )} / \sum{E_{total,i}}$\\
$return(E_{loss}, P_{aggr}$)
}

$\,$
\label{algo2}
\caption{Evaluate RL model  pseudocode.}
\end{algorithm}

Algorithm \ref{algo} shows the steps  to learn the optimal charging parameters $T_{BoostDur}$ and $P_{optimal}$. 
At each iteration, LearnRLModel()  function generates candidate values for charge session duration  $T_{maxboost}$ and low-power charge rate coefficient $P_{rate}$, applies them retrospectively on past historical data using EvaluateRLModel() (Algorithm \ref{algo2}) to compute the reward $R$ which depends on total energy loss $E_{loss}$   and the aggregate charging rate $P_{aggr}$. 
The candidate values that correspond to the highest value of the reward function are selected as optimal and are applied towards the next charging session. 
 Thus, the agent learns the parameters retrospectively, through trial and error using  historical data. 

The optimal policy search is accomplished using gradient descent with variable step size  \cite{NoceWrig06}. 
The experiments showed that the reward function is not concave, therefore variable step size allows to avoid getting stuck in a local minimum. 
The number of steps $n_{tries}$ was selected as 200 and the initial value of parameter boost duration $T_{BoostDur, i}$ was initialized to  an average session duration, which seemed to  perform well in the experiments.

\section{Experimental Results and Discussion}
\subsection{EV chargepoint dataset}
\label{secDataset}
The experimental evaluation is based on chargepoint usage datasets in the UK, provided by the Department of Transport Energy and Environment \cite{domesticdata2018}. 
The dataset contains raw data on the  amount of energy supplied and plugin duration per charging event. 
The dataset contains 3.17 million charging sessions from 25,126 domestic chargers collected in 2017 in the  format described in Table \ref{tab:domestic-format}. 
The charge points were funded by the UK government with the condition that the participants had to share their data for one year. 

\begin{table}[!h]
\caption{Domestic chargepoint dataset format and sample records.}
\label{tab:domestic-format}
\tiny
\setlength{\tabcolsep}{5pt}
\begin{tabular}{llllllll}
\hline
EventID	& CPID 		& StartDate	& StartTime	& EndDate	& EndTime	& Energy	& Duration\\
3177742	&	AN21771	&31/12/2017	&23:59:23		&01/01/2018	&18:20:23		& 8.8		&18.35\\
16679268	&	AN04715	&31/12/2017	&23:59:00		&01/01/2018	&00:03:00		&10.2	&0.066\\
16678965&	AN04849	&31/12/2017	&23:59:00		&01/01/2018	&13:40:00		&6.2		&13.68\\
3177556	&	AN21305	&31/12/2017	&23:57:23		&01/01/2018	&06:30:23		&9.5		&6.55\\
...
\end{tabular}
\end{table}

The charging event duration captures the duration of time the vehicle remains connected to the charge point and can be longer than the actual charging duration. 
The  session charging speed, computed separately by dividing the dispensed energy by the plugin duration, varies within the same charge point as 
plugin duration can be longer than the actual charging duration. 
To eliminate this factor from the analysis a novel  \textit{effective charging duration} metric was introduced,  defined as the ratio of dispensed energy to the maximum charging speed within the given EV charge point.  

 \begin{figure}[t]
 \centering
  \includegraphics[width=8.5cm,height=5.5cm]{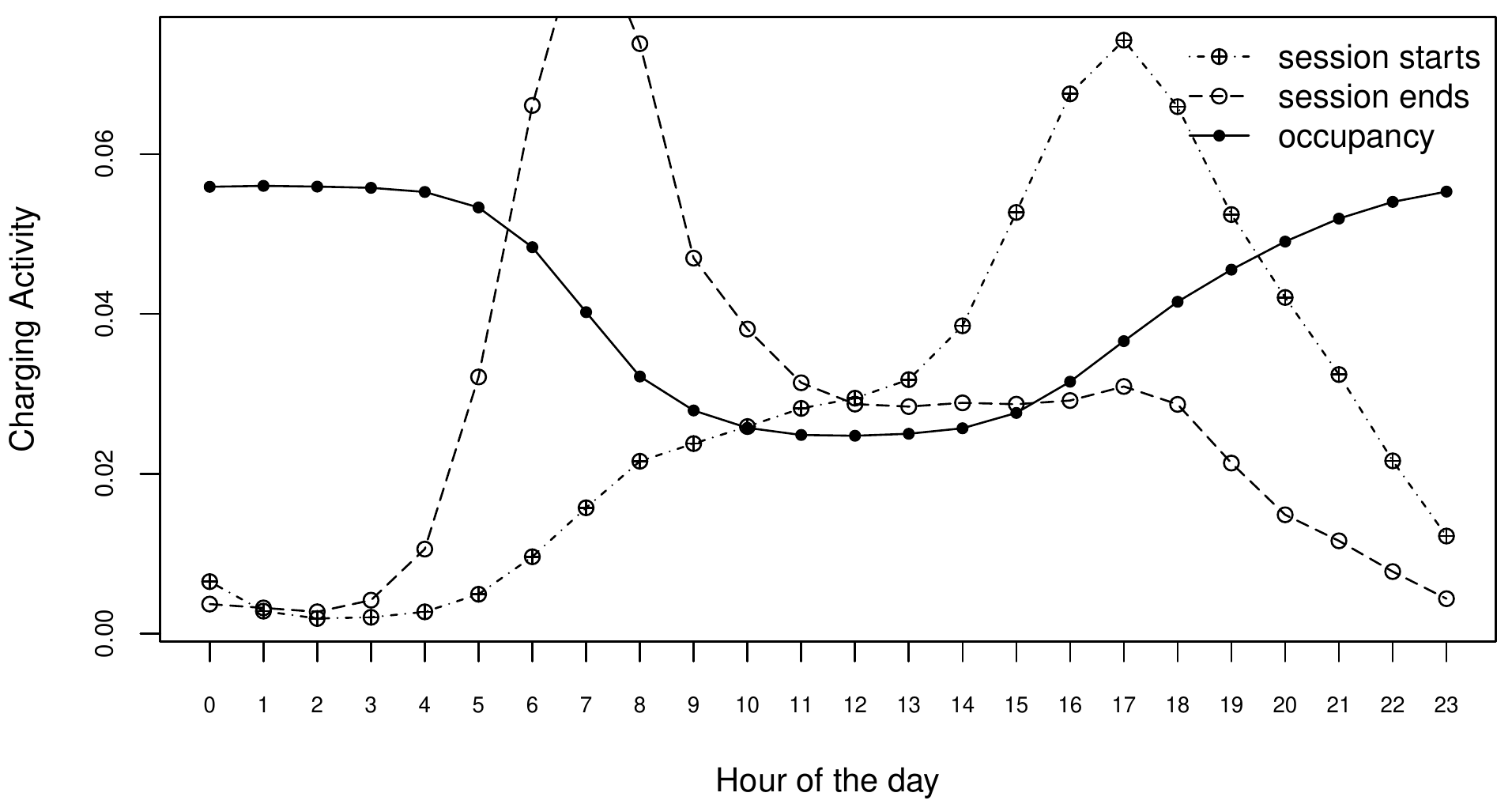}
  \caption{Domestic chargers temporal activity distribution. The sessions are typically initiated between  1\,pm and 9\,pm (dotted line). The distribution of time the EV remains connected  (solid line) is different due to the fact that most sessions are relatively long in duration. The mean and median plugin durations are 12.44 and 10.72 hours respectively. 
} 
  \label{fig:domestic-sum-by-hour}
\end{figure}

As domestic chargers can typically charge only one EV at a time, the overlapping sessions within the same charge point, which represented 1.7\% of all charging events have been removed from the analysis. 
Similarly, the charging events that are longer than 48 hours, representing approximately 1.8\% of the records, have been removed from the analysis as anomalous. 
Such events may appear due to a driver not closing the charging connector properly after the charging, which results in the session to continue to be recorded until a new charging event starts.  
More details about the dataset including information about data collection protocol, statistics and limitations are available at \cite{domesticdata2018}. 

Figure \ref{fig:domestic-sum-by-hour} shows that the charging session activity exhibits a strong temporal pattern. 
The charging sessions are typically initiated between  1\,pm and 9\,pm with  vehicles remaining connected until early in the morning. 
As most sessions are relatively long in duration (the mean and median plugin durations are 12.44 and 10.72 hours respectively), the distribution  EV charger occupancy is more spread throughout a day compared to the distribution of  session start times.
There are two major patterns in charging behavior. The first category of users plugin their vehicles in the evening and leave them connected in the morning.
The second category of usage pattern involves relatively shorter charging sessions throughout the day. 
The dataset has been used in a prior study  to  show how the peak demand can be reduced using battery-assisted charging systems that  accumulate energy during relatively quiet periods and release it during the peak hours to shorten the charge duration or reduce the peak load on the grid  \cite{ali2020isgt}.
In contrast, this work focuses on how to reduce peak demand by adaptively changing the charging rate without relying on a battery-assisted system.

\subsection{Results}
\label{secResults}

The main purpose of the experiments is to evaluate the performance of the adaptive charging and its impact  on the  individual and aggregate  reduction  of peak power usage. 
One of the goals is to understand the impact of history size on the algorithm performance, and whether storing an entire history is required.

The performance is measured in terms of the impact on the aggregate daily energy usage profile and the total energy deficit. 
The overall aggregate daily charging profiles were obtained by computing  the amount of dispensed energy in each daily timeslot  across all sessions for all charge points. 
The timeslot duration was set to 1-second to prevent the error accumulation due to energy quantization in each timeslot. 
For comparison, 60-second timeslot results in a significant discrepancy in total dispensed energy computation even between raw and hypothetical strategies.  
The energy deficit is defined as the difference between the target and the actual amount of dispensed energy and is always zero by definition for raw and hypothetical strategies.

The evaluation has been conducted for all charge points in the dataset with at least 10 charging sessions, which represent 22,731 charge stations. 
The training of the adaptive strategy was done on the first 80\% of sessions and testing on the last 20\% of the sessions, separately for each charge point and the results have been aggregated across all charge points. 
The gradient descent included 200 steps. 
The data analysis has been implemented and evaluated in $R$  package \cite{R}.

Finally, the  raw and hypothetical (ideal) charging strategies have been used as baselines for comparison. 
The raw charging strategy  assumes charging at the maximum possible rate until the target energy is met and then staying connected in idle mode. 
The ideal charging strategy uses a hypothetical scenario, where a perfect knowledge of each charging session duration is available, and the target energy is spread evenly throughout each session.

\subsection{The impact of history size}

\begin{figure}[t]
 \centering
  \includegraphics[width=9cm]{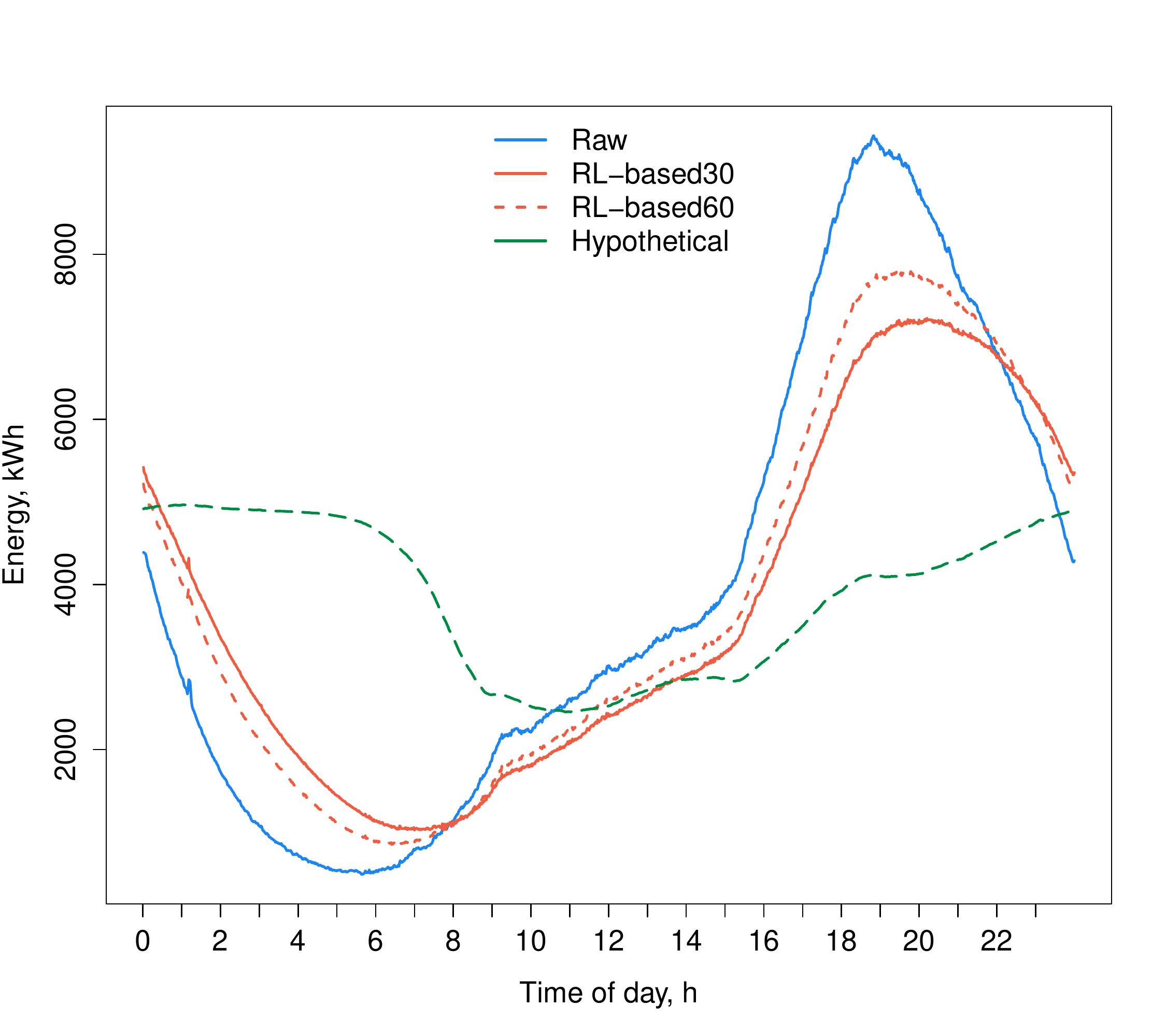}
  \caption{Adaptive learning performance aggregated across  all charge stations. The RL-based strategy with the history size of 30 results in 31\% peak reduction. }
  \label{fig:energyProfile22k}
\end{figure}

Figure \ref{fig:energyProfile22k} shows the aggregate energy profiles for different history sizes, as well as a comparison with those of  raw and hypothetical charging. 
As can be seen, the RL-based strategy reduces the peak by flattening the load and increasing the consumption during the nighttime. 
The performance of the reinforcement-learning based strategy depends significantly on the history size. 
Shorter history sizes reduce the peak power usage more aggressively but also result in higher energy deficits. 
For a history size of 30, the peak power usage reduces by as much as 31\% in the evening period compared to raw charging.  
As a comparison, a recent case study in Finland based on real data from 25,000 charging sessions collected over 2 years  from 8 charging sites shows that the peak loads at charging sites can be reduced by up to 55\% \cite{Fenner20peakshaving}.
However, the optimization strategy used in the study  computes  the peak load  as a ratio of dispensed energy to plugin duration, which requires the knowledge of the latter and corresponds to the hypothetical scenario in the presented study. 
\cite{zhang2018smartcharging} shows up to 80\% of peak reductions, however, it does not attempt to reduce the quality of charging service.

The  adaptive strategy may result in some energy deficit due to lower charging speeds in the slow charging phase.
The total energy deficit was 276,227\, kWh or just 5.0\% of total dispensed energy. 
Further analysis shows that 16\% of charge points have an energy deficit above 10\% of the total dispensed energy in that charge station. 
As history gets longer, the algorithm becomes more conservative as it evaluates the charging parameters over a wider range of drivers' behavior. 
For a history size of 60, the aggregate peak power reduces to 21\% with the total energy deficit reduced to 2.8\% or 159,934 kWh. 
The percentage of charge points with an energy deficit above 10\% is 8.9\%. 
Finally, for unlimited history size, the aggregate peak power reduced by 12.3\% with a total energy deficit of only 1.4\% or 78,882\,kW.  
The percentage of charge points with an energy deficit above 10\% reduces to 7.2\%.

\begin{figure}[t]
 \centering
  \includegraphics[width=8cm]{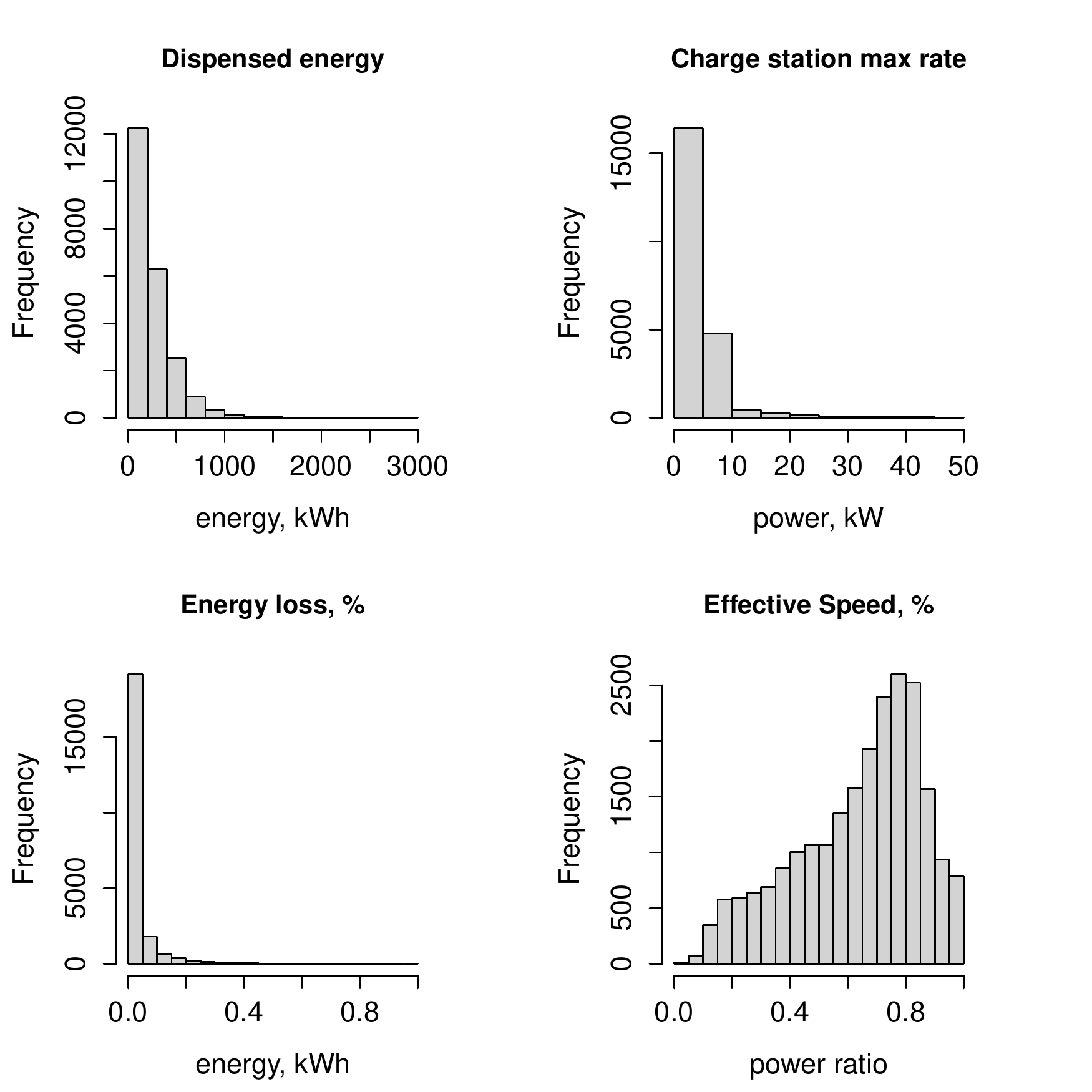}
  \caption{a) The distribution of target dispensed energy. b) The distribution of charge station max rates. c) The distribution of energy loss. d) The distribution of boost phase duration relative to raw charge duration. }
  \label{fig:4plots}
\end{figure}

\begin{figure}[t]
 \centering
  \includegraphics[width=9cm]{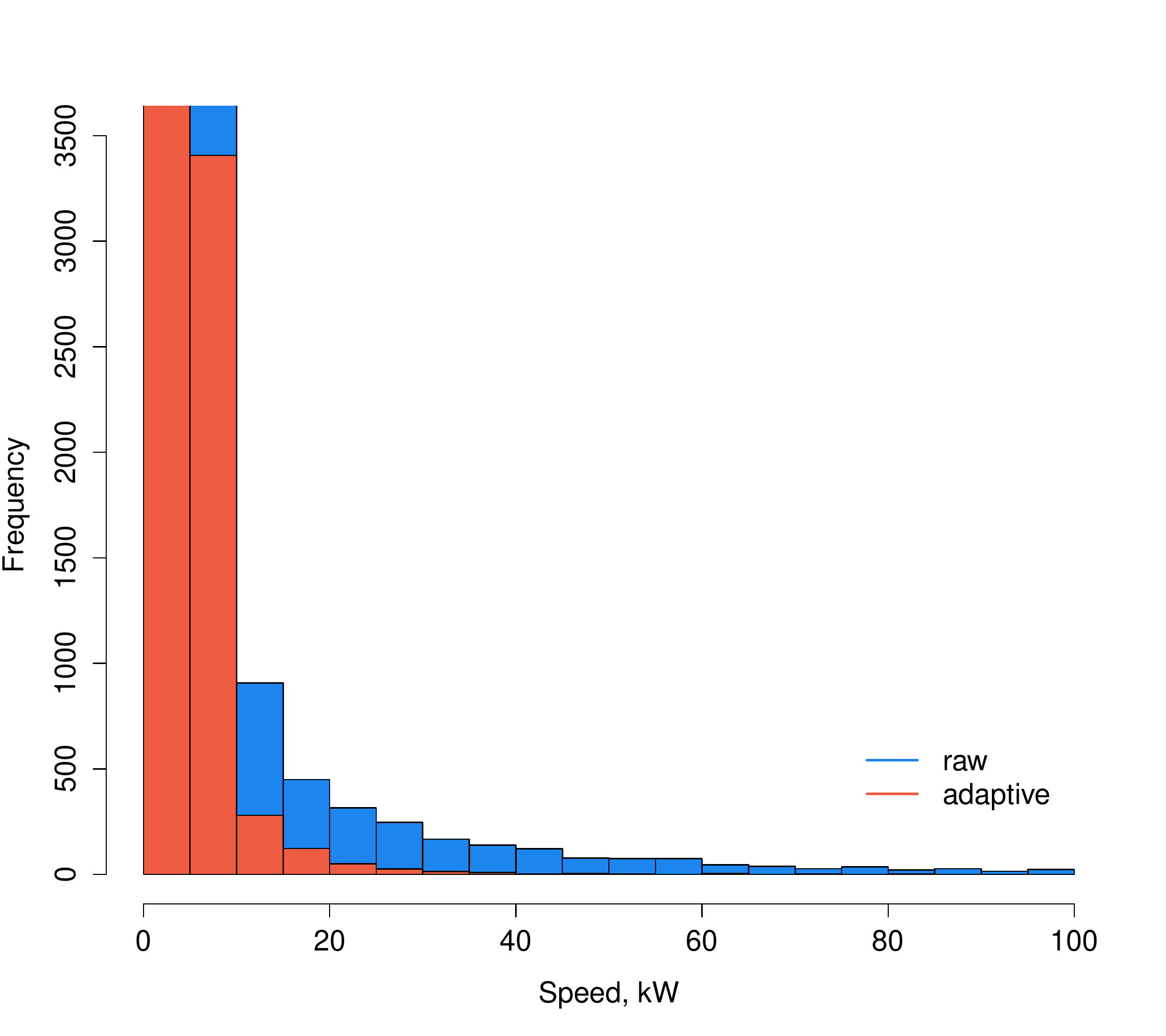}
  \caption{Comparison of charging speeds for raw and RL-based charging. The latter charging strategy results in significantly lower charging speeds. }
  \label{fig:speedDistribution}
\end{figure}

Figures \ref{fig:4plots}a-b show the distributions of total dispensed energy and maximum power rates for each charge station. 
Figure \ref{fig:4plots}c shows that a vast majority of sessions result in a very small energy loss relative to the total energy dispensed by the relative charge station. 
It can also be seen that a significant proportion of sessions charge at a rate much lower than the maximum power rate with a peak at 80\%, Figure \ref{fig:4plots}d. 
Figure \ref{fig:speedDistribution} compares  charging speeds for raw and RL-based charging. 
It can be seen that the adaptive  charging strategy results in significantly lower charging speeds.
The reduction in peak power usage in all reinforcement learning algorithm configurations is  lower than approximately 50\% reduction provided by the hypothetical strategy, Figure \ref{fig:energyProfile22k} (green line). 
However, it is important to note that  the adaptive algorithm does not require actual knowledge of the session duration as it directly controls the charging parameters that maximize the reward. 

\subsection{The impact on charging duration}
 \begin{figure}[t]
 \centering
  \includegraphics[width=9cm]{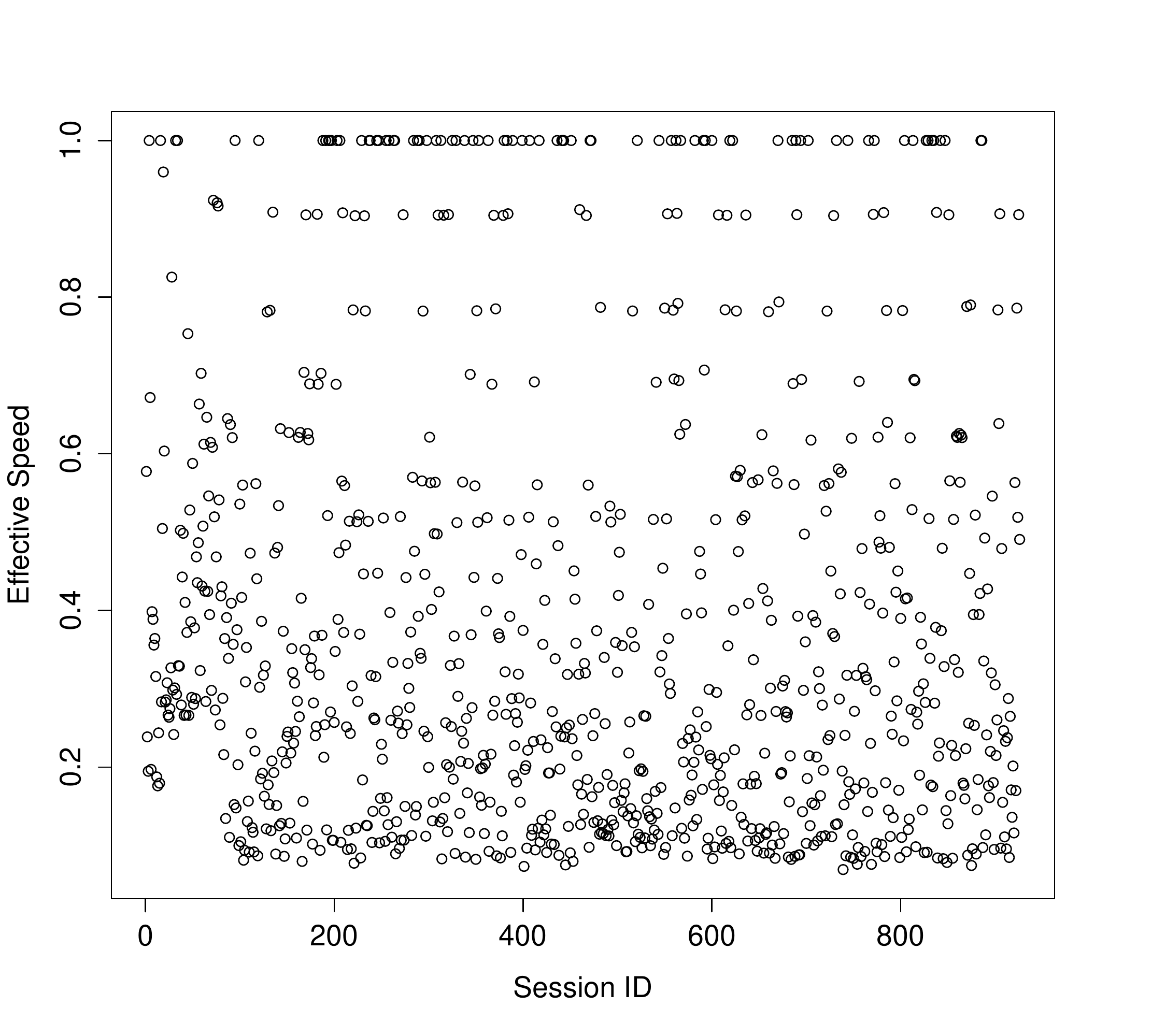}
  \caption{Effective speed distribution for CP AN15123. The mean and median session power rates are  37.8\% and 28.3\% of charge point capacity respectively.} 
  \label{fig:onlineLearningEffSpeed}
\end{figure}
 \begin{figure}[t]
 \centering
  \includegraphics[width=9cm]{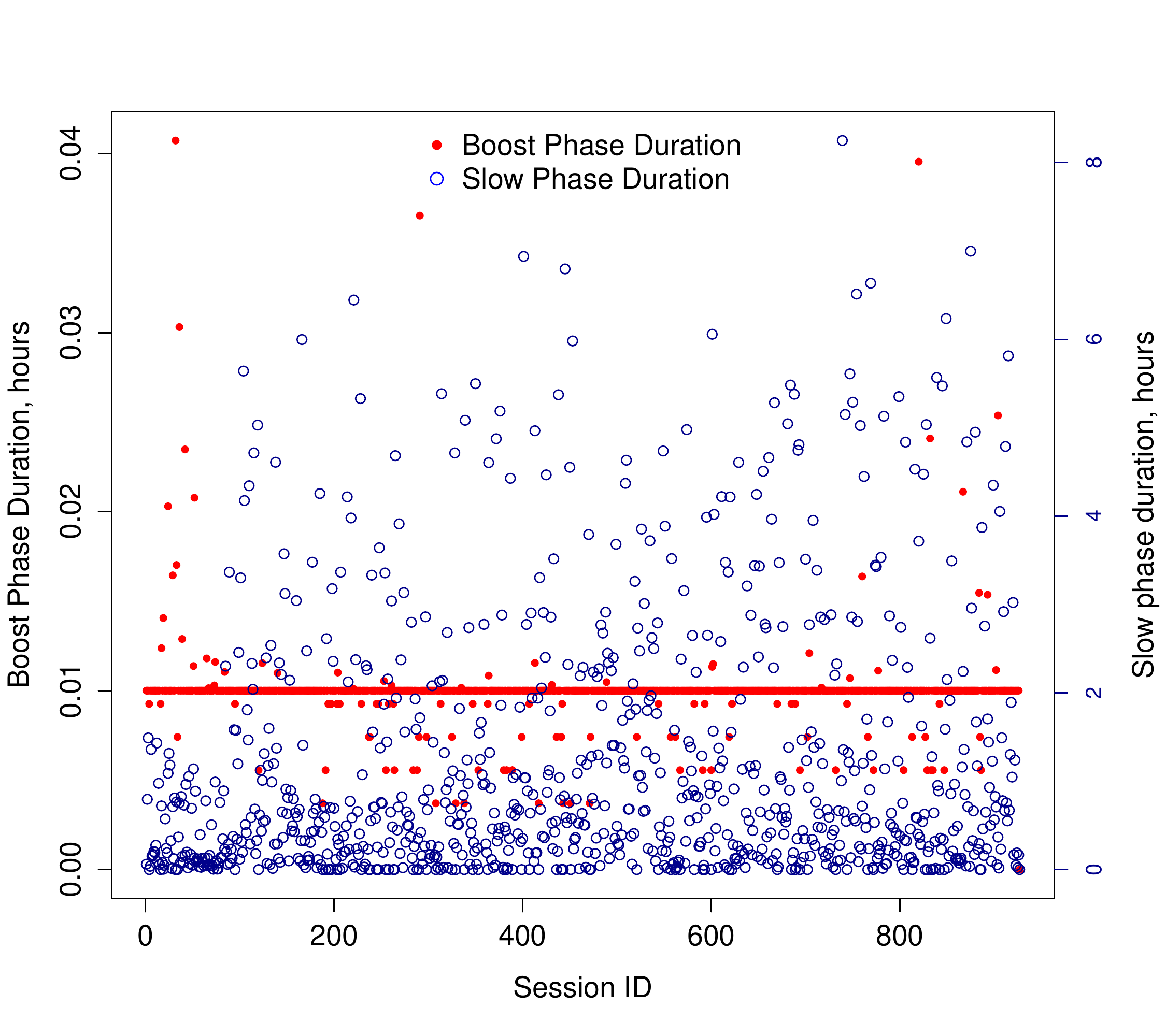}
  \caption{The distribution of boost phase and slow charge durations for CP AN15123. The boost phase (left axis) is typically much shorter than a slow phase duration (right axis). A large number of sessions have a boost duration of 0.01 hours. 
}
  \label{fig:onlineLearningBoostDur}
\end{figure}

\begin{table}[b]
\smaller
\setlength{\tabcolsep}{3pt}
\centering
\caption{Boost and slow charge phase  durations}
\begin{tabular}{ | c | c | c | c  | c | c | c |} 
 \hline
\multicolumn{2}{|c|}{RL30}  & \multicolumn{2}{c|}{RL60} & \multicolumn{2}{c|}{RL-ALL} & \multicolumn{1}{c|}{Raw} \\ 
 \hline
 BoostDur & SlowDur & BoostDur & SlowDur & BoostDur & SlowDur & EffectiveDur\\ 
 \hline
 0.30 & 3.47 & 0.40 & 2.92 & 0.49 & 2.55 & 1.82 \\ 
 \hline
\end{tabular}
\label{tab:BoostDur}
\end{table}

 \begin{figure}[t]
 \centering
  \includegraphics[width=9cm]{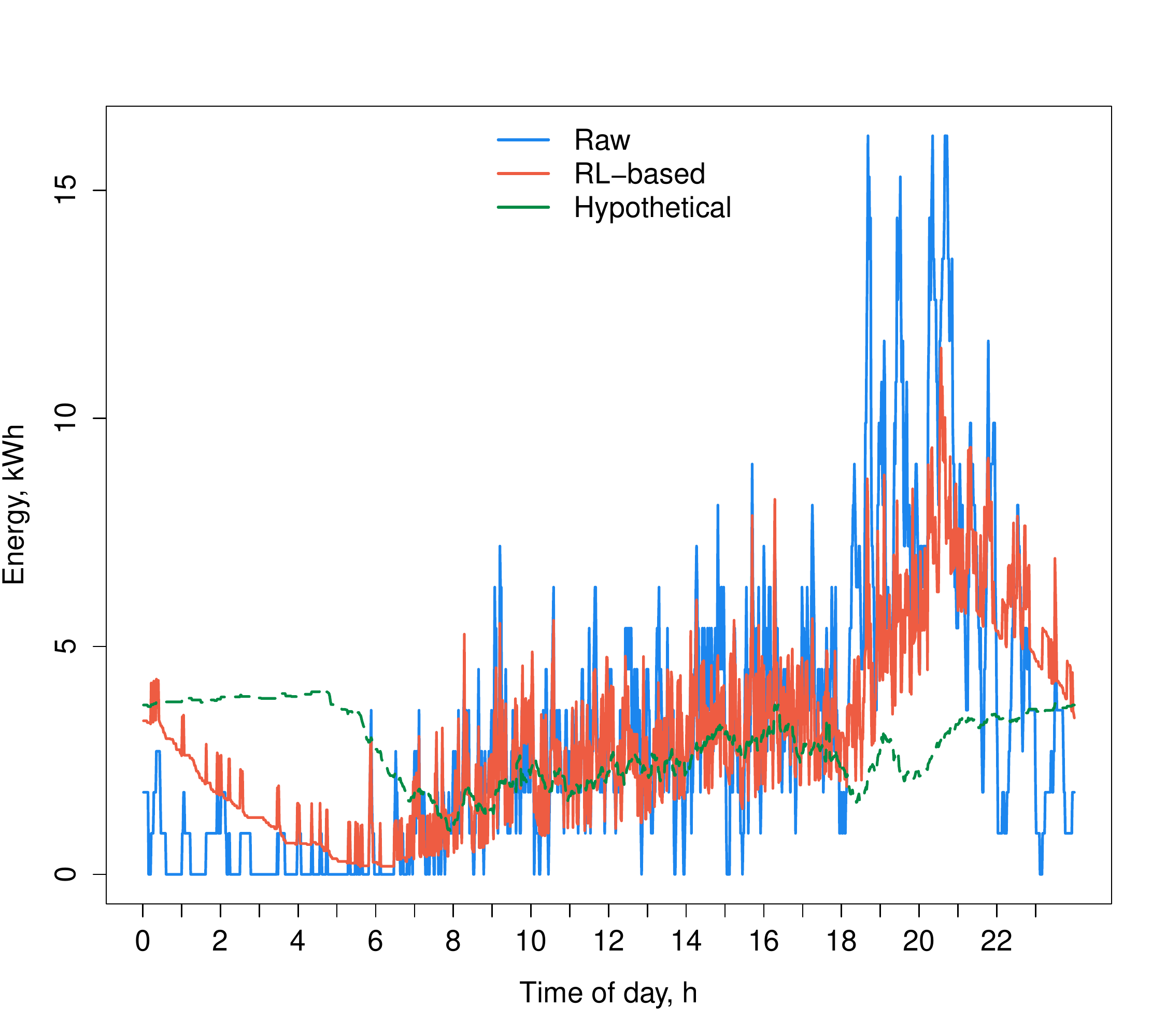}
  \caption{Online learning performance for CP AN15123. History size = 60. Reinforcement based strategy effectively shaves the peaks spreading the load in time. }
 \label{fig:onlineLearningProfile}
\end{figure}

The boost phase duration reduces the risk of undercharging as it charges the vehicle at the maximum power rate. 
In the case of battery-assisted charge stations, where a large capacity battery is used to accumulate energy in between charging sessions to boost charge the EV, the knowledge of boost charge duration can also be useful to estimate the required charge point battery capacity or required charge level. 

Table \ref{tab:BoostDur} compares the boost and slow charge phase durations for all three strategies. 
In all cases, the boost phase duration is much shorter than an average effective charging rate of 1.82 under raw charging. 
The latter is computed as the ratio of dispensed energy by the point power rate event and averaged across all sessions in all charge stations. 
It can be seen that as the history sizes increases, the algorithm gets more conservative and allocates more time for boost charging.

\subsection{Online learning case study}
This section describes an evaluation of the approach in an online learning mode, where the agent optimizes  charging parameters after each  charging session, similar to how it would operate in a real deployment. 
The performance is illustrated on a charge point AN15123, selected because it was the busiest charge point in the dataset with a charging rate above 7\,kW.

The charge point has a maximum charging rate of 54\,kW, and contained  1031 sessions. 
Figures \ref{fig:onlineLearningEffSpeed}, \ref{fig:onlineLearningBoostDur}, \ref{fig:onlineLearningProfile} show the  performance of the adaptive approach on one specific charge point selected arbitrarily. 
The learning starts after the first 100 sessions, which represents  10\% of all charging sessions for all sessions for this charge point.

Figure \ref{fig:onlineLearningEffSpeed} shows the effective speed for the proposed strategy. 
The median and mean effective speed ranges are only 0.28 and 0.37  of the maximum charging speed respectively. 
The total energy deficit for the adaptive strategy was just  54\,kWh, which represents 1.3\% of all the energy dispensed by the charge point. 
The algorithm effectively tracks user behavior and adapts the charging parameters accordingly.

Figure \ref{fig:onlineLearningBoostDur}  compares the boost and slow phase durations for all sessions. 
It can be seen that boost phase duration is typically small and below 0.12 hours, whereas slow charge durations can last up to 8.65 hours. 
The mean effective charging speed, boost, and slow phases are 18.29\,kW, 0.02 hours and 1.19 hours respectively. 
Figure \ref{fig:onlineLearningProfile} compares the overall energy profiles for raw, adaptive, and ideal charging strategies. 
The raw charging results in sharp peaks in power consumption with the highest peak at around 6-9\,pm.  
The adaptive charging strategy visibly reduces the peaks spreading the load in time. 
The hypothetical strategy results in the highest peak reduction as it spreads the power evenly throughout each session. 
However, it should be noted that it requires a perfect knowledge of plugin session duration, which is difficult to predict in practice, as was shown in the previous section.

\subsection{Discussion}
The study focuses on the maximum potential for reducing peak charging demand for individual charging stations using local historical information only. 
The algorithm requires storing the past charging session history in its memory. 
Since each session requires the storage of 3 values (start timestamp, end timestamp, dispensed energy). 
Assuming 4 bytes  for the first two and 2 bytes each for the latter, the annual data requirement will be approximately 3,650 bytes if the charging happens once a day. 
The busiest domestic charge point in the dataset contained 1,381 sessions, which can be stored in just 13,810 bytes. 
The algorithm should be suitable for  implementation in an embedded  platform and does not have significant computational overhead. 

In this study, the proposed charging profile is similar to a step function, consisting of  discrete high and low-speed phases. 
The learning algorithm searches for the parameters of step height and step duration. 
 However, it may be possible to define a more general charging profile that takes into account battery characteristics, health, and other factors. 
The more general approach would need to optimize the parameters of this function. 
Investigating these ideas is a potential future work. 
This research focuses on domestic charging, where  there is a significant potential for energy coordination.
Public charge points as data indicates are characterized by frequent and short sessions, which are likely made at high speed.

The algorithm requires the knowledge of the maximum charging rate $P_{max}$, which is limited by both  EV and charge point  capabilities. 
In  the experiments, the value of $P_{max}$ was selected as the charger point's maximum charging power throughout the entire year. 
This assumes that each household has a single electric vehicle, which should be a reasonable assumption considering today's price of  EVs.

\section{Conclusions and Future Work}
\label{secConclusion}

This paper  proposes a novel approach for smart electric vehicle charging that  identifies optimal charging parameters by training on a history of past charging sessions using reinforcement learning. 
Unlike other approaches, the proposed reinforcement learning algorithm does not require predicting users' behavior and learns through trial and error through analysis of past behavior. 
The proposed algorithm can be used in situations, where a user needs to minimize peak current without modifying his or her behavior while not using any coordination with the grid. 
The optimal boost charge duration learned by the algorithm can also be useful in household energy scheduling to estimate a target storage battery level for electric vehicle charging. 

The evaluation shows the extent to which the approach can reduce aggregate peak current if used collectively by several thousand charge points across the UK.  
The approach is not computationally intensive and can be implemented on relatively low-cost hardware. 
In this work, the reinforcement learning agent does not take into account such factors as the day of the week, time of day, weather, or other contextual information. 
It should be possible to extend the proposed approach to take into account additional factors, which is a potential future work. 

\bibliographystyle{IEEEtran}
\smaller
\bibliography{references}
\end{document}